\documentclass[a4paper,11pt]{article}
\pdfoutput=1 
\usepackage{jcappub} 
\title{\boldmath Short-baseline electron antineutrino disappearance study
by using neutrino sources from $^{13}$C + $^{9}$Be reaction}

\author[a]{Jae Won Shin,}
\author[a,1]{Myung-Ki Cheoun,\note{Corresponding author.}}
\author[b,c]{Toshitaka Kajino }
\author[d]{and Takehito Hayakawa}

\affiliation[a]{Department of Physics and Origin of Matter and Galaxy Evolution (OMEG) Institute,
Soongsil University,\\Seoul 156-743, Korea}
\affiliation[b]{Division of Theoretical Astronomy,
National Astronomical Observatory
of Japan,\\Mitaka, Tokyo 181-8588, Japan}
\affiliation[c]{Department of Astronomy, Graduate School of Science, University of
Tokyo,\\Hongo, Bunkyo-ku, Tokyo 113-0033, Japan}
\affiliation[d]{Quantum Beam Science Directorate (QUBS),
Japan Atomic Energy Agency (JAEA),\\2-4 Shirane, Shirakata, Tokai-mura, Naka-gun, Ibaraki 319-1195, Japan}

\emailAdd{shine8199@skku.edu}
\emailAdd{cheoun@ssu.ac.kr}
\emailAdd{kajino@nao.ac.jp}
\emailAdd{hayakawa.takehito@qst.go.jp}

\date{Feb. 26, 2017 (v-4)}

\abstract{
To investigate the existence of sterile neutrino,
we propose a new neutrino production method
using $^{13}$C beams and a $^{9}$Be target
for short-baseline electron antineutrino
(${\bar{\nu}}_{e}$) disappearance study.
The production of secondary unstable isotopes which can emit neutrinos
from the $^{13}$C + $^{9}$Be reaction is calculated
with three different nucleus-nucleus (AA) reaction models.
Different isotope yields are obtained using these models,
but the results of the neutrino flux
are found to have unanimous similarities.
This feature gives an opportunity to study neutrino oscillation through shape analysis.
In this work, expected neutrino flux and event rates are discussed in detail
through intensive simulation of the light ion collision reaction
and the neutrino flux from the beta decay
of unstable isotopes followed by this collision.
Together with the reactor and accelerator anomalies,
the present proposed ${\bar{\nu}}_{e}$ source is shown to be
a practically alternative test of the existence
of the $\Delta m^{2}$ $\sim$ 1 eV$^{2}$ scale sterile neutrino.
}

\begin{document}
\maketitle
\flushbottom

\section{Introduction}

Over the past few decades,
a considerable number of studies has been done on the neutrino oscillation
with a great success of measuring neutrino mixing angles.
However, some experiments for the neutrino oscillation revealed more
or less disagreements with the three-flavor neutrino model,
which termed as neutrino anomalies,
as reported in LSND\,\cite{LSND_1}, MiniBoone\,\cite{MiniB_1}, reactor experiments\,\cite{reacAnt}
and gallium experiments\,\cite{GaAnomaly}.
One of the approaches for explaining the neutrino anomalies
is to presume the existence of the hypothetical fourth neutrino,
which is called as sterile neutrino,
because the sterile neutrino does not interact
with other particles except for a mixing with active neutrinos.

The possibility of the existence of sterile neutrino triggered
by the lots of anomalies is now being widely discussed in the filed of particle,
nuclear physics, and astrophysics including cosmology.
Most of the neutrino experiments are classified into the long-base line accelerator
and the short-base line reactor neutrino experiments.
The former neutrino source comes from high-energy proton accelerators;
this source produces high-energy and high-flux neutrinos
from the pion or kaon decay at rest or in flight and utilize long-distance detectors
from the accelerator
except for the MiniBooNE-like experiments.
On the contrary, the latter one relies on neutrinos from nuclear reactors,
which enables to use relatively the short-distance detector from the source,
but needs to pin down the ambiguity on the neutrino spectrum stemming
from vast numbers of nuclear fissions in the reactor \cite{Baha16}.
Here, it should be noted that recent results from the IceCube neutrino telescope,
which limits on the mixing of sterile neutrino and muon neutrino,
does not constrain the neutrino mixing angle relevant to the reactor anomaly \cite{IceCube16}.

Recently many interesting studies for the existence of sterile neutrinos (${\nu}_{s}$)
are proposed
with antineutrino sources from radioactive isotopes
or an accelerator-based
Isotope Decay-At-Rest (IsoDAR) concept
\cite{Ce144s1,Ce144s2,Ce144s3,sterileNu1,isoDar1}.
They utilize different neutrino sources from the two neutrino sources mentioned above.
For instance, KamLAND (CeLAND) \cite{Ce144s2} and
Borexino (Short distance neutrino Oscillations
with BoreXino (SOX)) \cite{Ce144s3} plan to perform experiments
using approximately 100 kCi of $^{144}$Ce-$^{144}$Pr
radio isotopes which can generate
antineutrinos with energy of up to 3 MeV.
As another type source,
electron antineutrinos (${\bar{\nu}}_{e}$)
produced by $^{8}$Li using an accelerator-based IsoDAR concept
was proposed \cite{sterileNu1, annu8Li1}.
The ${\bar{\nu}}_{e}$ from $^{8}$Li has higher energy than that from
$^{144}$Ce-$^{144}$Pr,
and thus can be used for study of
antineutrino spectrum distortion
in the energy region of 5 MeV $< E_{\bar{\nu}} <$ 7 MeV,
where
some distortions or anomalies are reported by
the reactor antineutrino experiments
(Daya Bay \cite{ReactBump_Day}, Double Chooz \cite{ReactBump_DOU1}
and RENO \cite{ReactBump_RENO1, ReactBump_RENO2}).

On the other hand, there are many ion beam facilities
such as FAIR at GSI, FRIB at MSU,
HRIBF at ORNL, ISAC at TRIUMF, ISOLDE at CERN,
RIBF at RIKEN,
SPES at INFN, and SPIRAL2 at GANIL.
They have been designed
or are being developed to obtain features
(e.g. use of high-intensity high-power primary beams,
use of large-aperture high-field superconducting magnets, etc.)
suitable for variety scientific purposes \cite{plan1, plan2, plan3, plan4}.
The ion beams can provide us with
new opportunities in neutrino physics,
especially, for production of artificial neutrino sources.
By using the ion beams with compact neutrino detectors such as
DANSS \cite{danss},
NEUTRINO4 \cite{neu_4},
NUCIFER \cite{nuci},
PANDA \cite{panda},
PROSPECT \cite{prospect_1},
and STEREO \cite{stereo} which have been planned to measure reactor neutrinos
at a distance of several meters,
we can test the existence of sterile neutrinos,
in particular, on 1 eV mass scale.

In this work, to study the possibility of sterile neutrino, we propose a new neutrino
production method
with a $^{13}$C beam and a $^{9}$Be target.
Unstable isotopes such as $^{8}$Li and $^{12}$B can be
produced through the $^{13}$C + $^{9}$Be reaction and decay subsequently.
Thus, we can obtain new neutrino sources (${\bar{\nu}}_{^{13}C + ^{9}Be}$)
from the possible beta decay processes of
unstable isotopes produced at the $^{13}$C + $^{9}$Be reaction.
In a sense, the production mechanism is similar to
that of reactor neutrinos.
However, in this case, neutrino energy spectrum is much easier identified compared with those of reactor neutrinos
because the number of isotopes is limited and the geometry is simple.
The production of secondary isotopes from
the $^{13}$C + $^{9}$Be reaction is calculated using
the GEANT4
particle transport Monte Carlo code
\cite{g4n1, g4n2},
with three different nucleus-nucleus (AA) reaction models.
Different isotope yields are obtained using these models,
but the results of the neutrino flux
are shown to have a striking similarity.
This unique feature gives a chance
to neutrino oscillation study through shape analysis
regardless of the theoretical AA models considered.
Expected neutrino flux and event rates
including the sterile neutrino contribution are discussed in this work.

Outline of the paper is as follows.
In section \ref{meth-sec},
proposed experimental setup,
the neutrino detection and
simulation tools including the nuclear reaction model
are described.
In section \ref{result-sec},
the results of
expected neutrino flux
and event rates are presented,
and neutrino disappearance features and possible reaction rate changes
by the sterile neutrino are discussed with possible inherent errors in section \ref{sterile-sec}.
A summary is given in section \ref{sum-sec}.

\section{Methods
\label{meth-sec}}

\subsection{Proposed experimental setup
\label{meth-sec_setup}}

\begin{figure}[tbp] 
\centering
\epsfig{file=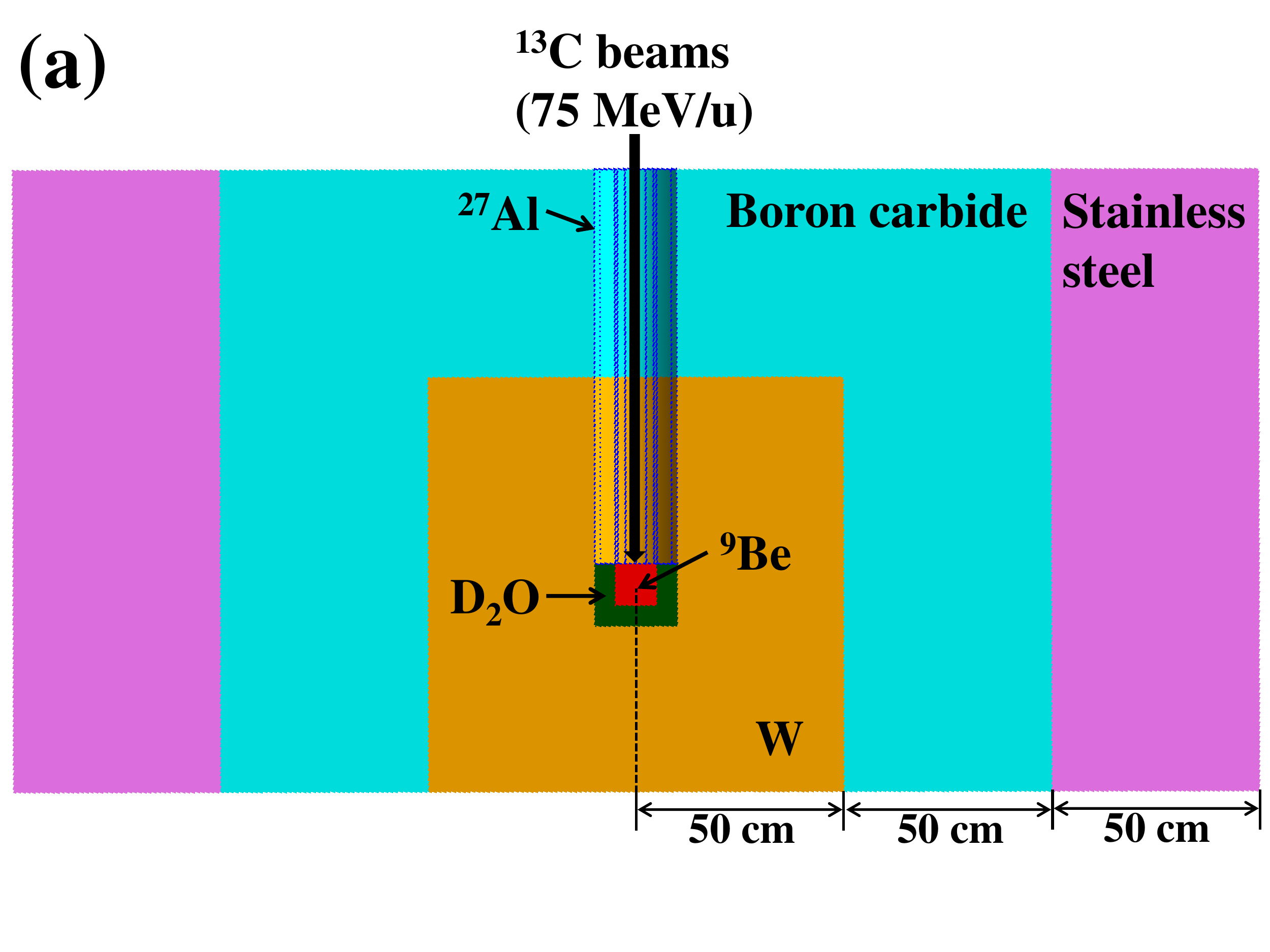, width=3.in}
\epsfig{file=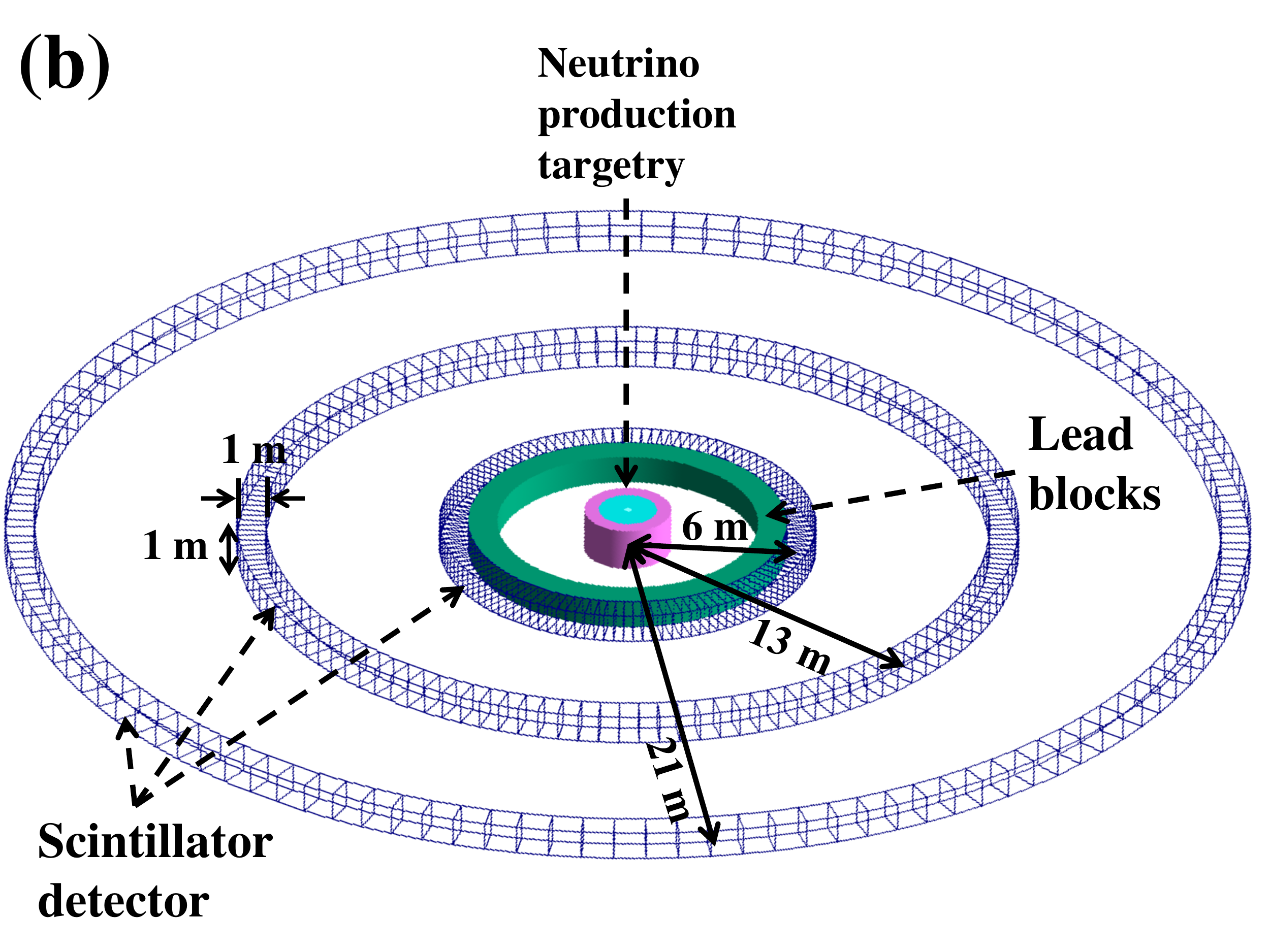, width=3.in}
\caption{(Color online)
(a) A schematic cross section view showing neutrino production targetry.
(b) A schematic view showing proposed short baseline experiment setup.
}
\label{fig1}
\end{figure}

We propose an electron antineutrino source using
an accelerator-based IsoDAR concept
with a $^{13}$C beam and a $^{9}$Be target.
75 MeV/u $^{13}$C beam with
a current of 300 p$\mu$A,
namely 293 kW of beam power, is considered in this work.
There are many ion beam facilities which have been designed
and are being developed with advanced features
(e.g. use of high-intensity high-power primary beams).
For example, a beam power of 400 kW
is aimed to achieve at FRIB (MSU).
In this regard,
the present proposal is a feasible experimental plan.

Figure \ref{fig1} (a) shows geometrical setups
for the neutrino production.
The $^{9}$Be target
is modeled as a cylinder of a radius of 5 cm
and a thickness of 10 cm.
Neutrinos
are obtained from
unstable isotopes produced through
the $^{13}$C + $^{9}$Be reaction.
The $^{9}$Be target is surrounded by
a D$_{2}$O layer with 5 cm thickness
for cooling.
In addition, tungsten, boron carbide, and stainless steel layers
surround the $^{9}$Be target and the D$_{2}$O layer
for effective secondary neutron generation and shielding of $\gamma$-rays.

The electron antineutrino source proposed in this work
can be useful for the neutrino disappearance study of
the investigation of the existence of fourth neutrino, sterile neutrino.
Figure \ref{fig1} (b) shows the proposed short-baseline experiment setup
consisting of
the neutrino production targetry
and the detection sector.
Lead blocks with 1 m thickness
surround the neutrino production targetry
to shield the radiation from the targetry.
In this work,
we consider three ring-shape detectors
with a height of 1 m and a thickness of 1 m.

\subsection{Electron antineutrino detection
\label{meth-sec_nuDetec}}

For electron antineutrino detection,
an inverse beta decay (IBD) reaction,
${\bar{\nu}}_{e} + p \to e^{+} + n$,
is considered in this work.
This IBD reaction in neutrino detection
offers two signals :
a prompt signal due to the annihilation of a positron
and a delayed signal of 2.2 MeV $\gamma$-ray following
neutron capture.
Characteristics of two distinct detections can
make efficient rejection
of possible backgrounds.
The energy dependent cross section of the IBD reactions
can be expressed by \cite{ibdCS}
\begin{eqnarray}
\sigma_{{\bar{\nu}}_{e}}(E_{{\bar{\nu}}_{e}})
  \approx p_{e} E_{e}
  E_{{\bar{\nu}}_{e}}^{-0.07056+0.02018{\rm{ln}}E_{{\bar{\nu}}_{e}} - 0.001953 {\rm{ln}}^{3} E_{{\bar{\nu}}_{e}}} \times 10^{-43} [\rm{cm}^{2}],
\label{eq:IBDcs}
\end{eqnarray}
where $p_{e}$, $E_{e}$, and $E_{{\bar{\nu}}_{e}}$ are
the positron momentum, the total energy of the positron,
and the energy of ${\bar{\nu}}_{e}$ in MeV, respectively.
Note that the mass difference between $m_{n}$ and $m_{p}$
can be described as
$E_{e} = E_{{\bar{\nu}}_{e}} - \Delta$, where
$\Delta = m_{n} - m_{p} \approx 1.293 ~\rm{MeV}$.
This cross section agrees within few per-mille
with the fully calculated results including
the radiative corrections and the final-state interactions in IBD.

\subsection{Simulation tool and nucleus-nucleus reaction models
\label{meth-sec_g4}}

We have performed
GEANT4 \cite{g4n1, g4n2} simulations
for estimating the production yields of isotopes
by considering the bombardment of 75 MeV/u $^{13}$C
beams on the $^{9}$Be target as shown in figure \ref{fig1}.
In these calculations, we use
the G4ComponentGGNuclNuclXsc class for the calculation of
overall cross sections of AA reactions and
the G4ionIonisation class for ionization processes.
The G4ComponentGGNuclNuclXsc class
provides total, inelastic, and elastic cross sections
for the AA reactions
using the Glauber model with Gribov corrections \cite{G4GG_1,G4GG_2,G4GG_3,G4GG_4}
where this class is valid for all incident energies above 100 keV.
The G4ionIonisation class is used for the calculation of the ionization processes,
where the effective charge approach \cite{g4_ioni} and
the ICRU 73 \cite{g4_dedx1} table for the stopping power are used.

An important point in the calculation of
heavy ion collision reactions is the choice
of hadronic models.
To discuss the relative isotope production yields caused by
different AA reactions,
three different hadronic models;
G4BinaryLightIonReaction \cite{BICref1, g4_bicLion},
G4QMDReaction \cite{g4qmdRef} and G4INCLXXInterface \cite{INCL_1, INCL_2}
are used.
They are described in detail in the Physics Reference Manual \cite{g4_physRef}
on the web \cite{g4_web}.
Here, we list some key features of them for the readers' convenience.
We distinguish our model calculations by referring to
G4BinaryLightIonReaction, G4INCLXXInterface, and G4QMDReaction simply as
``G4BIC", ``G4INCL" and ``G4QMD", respectively.

G4BIC :
The G4BinaryLightIonReaction class
is an extension of G4BinaryCascade for light ion reactions.
It is a data driven from the Intra-Nuclear Cascade model based on a detailed 3-D model of
the nucleus and binary scattering
between reaction participants and nucleons within the nuclear model.
Participant particles are either a primary particle including nucleons in a projectile nucleus,
or particles generated or scattered in the process of cascade.
Each participating particle
is seen as a Gaussian wave packet and total wave function is
assumed to be a direct product of the wave functions of the
participating particles without antisymmetrization.
The equations of motion has the same structure as the
classical Hamilton equations, where the Hamiltonian is calculated from a simple time-independent optical potential.
The nucleon distribution follows a Woods-Saxon model for heavy nuclei (A$>$16)
and a harmonic-oscillator shell model for light nuclei (A$<$17).
Participant particles are only propagated in the nucleus,
and participant-participant interactions are not taken into account in the model.
The cascade terminates
when mean kinetic energy of scattered particles (participants within the nucleus)
has dropped below a threshold (15 MeV).
After the cascade termination,
properties of the residual excitation system and the final nuclei are evaluated.
Then,
the residual participants and the nucleus
in that state are treated by pre-equilibrium decay.
For statistical description of particle emission from excited nuclei,
the G4ExcitationHandler class is used.
For light ion reactions, projectile excitations are determined
from the binary collision participants using the statistical approach towards
excitation energy calculation in an adiabatic abrasion process.
Given this excitation energy,
the projectile fragment is then treated by
the G4ExcitationHandler.

G4QMD :
G4QMDReaction is based on
a quantum extension of the classical molecular dynamics (QMD).
It is widely used to simulate AA reactions for many-body processes,
in particular, the formation of complex fragments.
QMD has similar characteristics to Binary Cascade
in treating each participant particle as a Gaussian wave packet and
assuming total wave function to the direct product of the participants.
Comparing with Binary Cascade, however,
QMD has some different characteristics such as
the definition of a participant particle,
potential term in the Hamiltonian
and participant-participant interactions.
Participant particles in the QMD mean
entire nucleons in the target and projectile nucleus.
The potential terms of the Hamiltonian in QMD
are calculated from the entire relation of particles in the system
where the potential includes a Skyrme type interaction,
a Coulomb interaction and a symmetry term.
Because there is no criterion between participant particle
and others in QMD,
participant-participant interactions are naturally included.
There are many different types of QMD models, but
G4QMDReaction is based on
JAERI Quantum Molecular Dynamics (JQMD) \cite{niita1}.
The self-generating potential field is used in G4QMD, and
the potential field and the field parameters of G4QMD
are also based on JQMD with Lorentz scalar modifications.
In the G4QMDReaction,
the reaction processes are also described by two steps.
First, as a dynamical process,
the direct reactions, non-equilibrium reactions, and dynamical
formation of highly excited fragments are calculated in the
manner of QMD.
Second, as a statistical process,
evaporation and fission decays
are performed for the excited nucleons produced in the
first step.
As the excitation model,
GEM (generalized evaporation model) \cite{GEM} is used.

G4INCL :
This class is being used for reactions induced
by nucleons, pions and light ion on any nucleus
using the INCL++ model, which is a version of
Li\`{e}ge intranuclear-cascade model (INCL) \cite{INCL_0}
fully re-written in C++.
For light ion induced reactions,
the projectile is described as a collection
of independent nucleons with gaussian momentum
and position distributions which use the realistic standard deviation of the projectile ion
for the position distribution.
Momenta and positions of the nucleons inside
a target nucleus are determined
by modeling the nucleus as a
free Fermi gas in a static potential well with a realistic density.
The reaction is described as an avalanche of binary nucleon-nucleon collisions,
which can lead to the emission of energetic particles and to the formation of an
excited thermalised nucleus as remnant.
Particles in the model are labeled either as participants (projectile particles
and particles that have undergone a collision with a projectile)
or spectators (target particles that have not undergone any collision).
Collisions between spectator particles are neglected.
The projectile (light ion) follows globally
a classical Coulomb trajectory until one
of its nucleon impinges on
a spherical calculation volume
around the target nucleus, which is large enough to
marginally neglect nuclear interactions.
The nucleons entering the calculation sphere
move globally (with the beam velocity) until one of them
interacts with a target nucleon.
The nucleon-nucleon (NN) interaction is then computed with the individual momenta with Pauli
blocking restriction.
Nucleons crossing the sphere of the calculation
without any NN interactions are also combined in
the ``projectile spectator" at the end of the cascade.
The cascade stops when the remnant nucleus shows signs
of thermalization;
a rather unique aspect of INCL is the
self-consistent determination of the cascade stopping time.
The projectile spectator nucleus is kinematically defined
by its nucleon content and its
excitation energy obtained
by an empirical particle-hole model, and then
the de-excitation of the projectile fragments
is described by the G4ExcitationHandler class.
%
%

Many different isotopes produced via
the $^{13}$C + $^{9}$Be reaction can emit neutrinos
with various energies.
The energy distributions of the neutrinos
are calculated using
``G4RadioactiveDecay" \cite{G4RDM_0, G4RDM_1} class
based on the Evaluated Nuclear Structure Data File (ENSDF) \cite{G4ensdf}.


\section{Results
\label{result-sec}}

\begin{figure}[tbp]
\centering
\epsfig{file=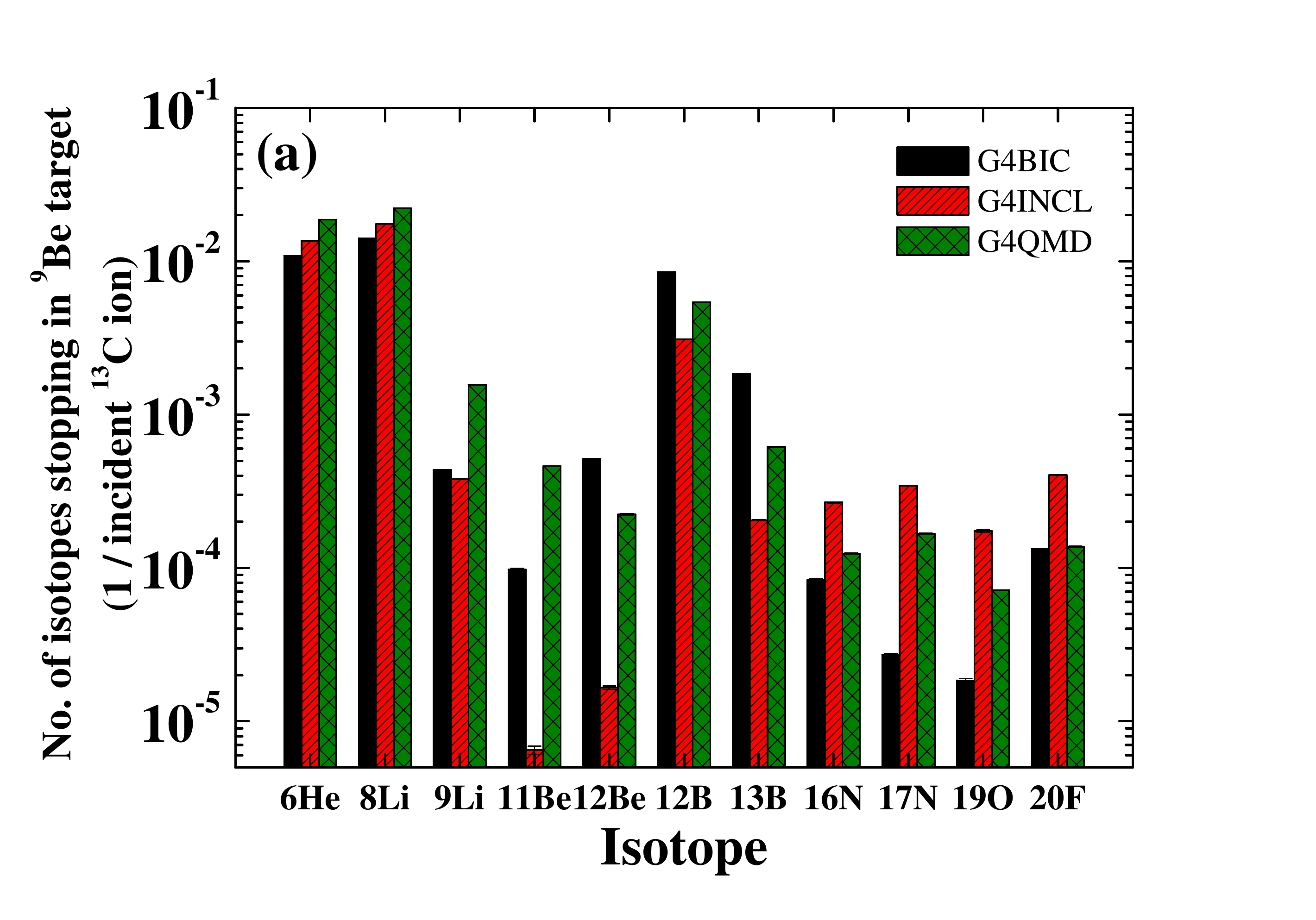, width=3.in}
\epsfig{file=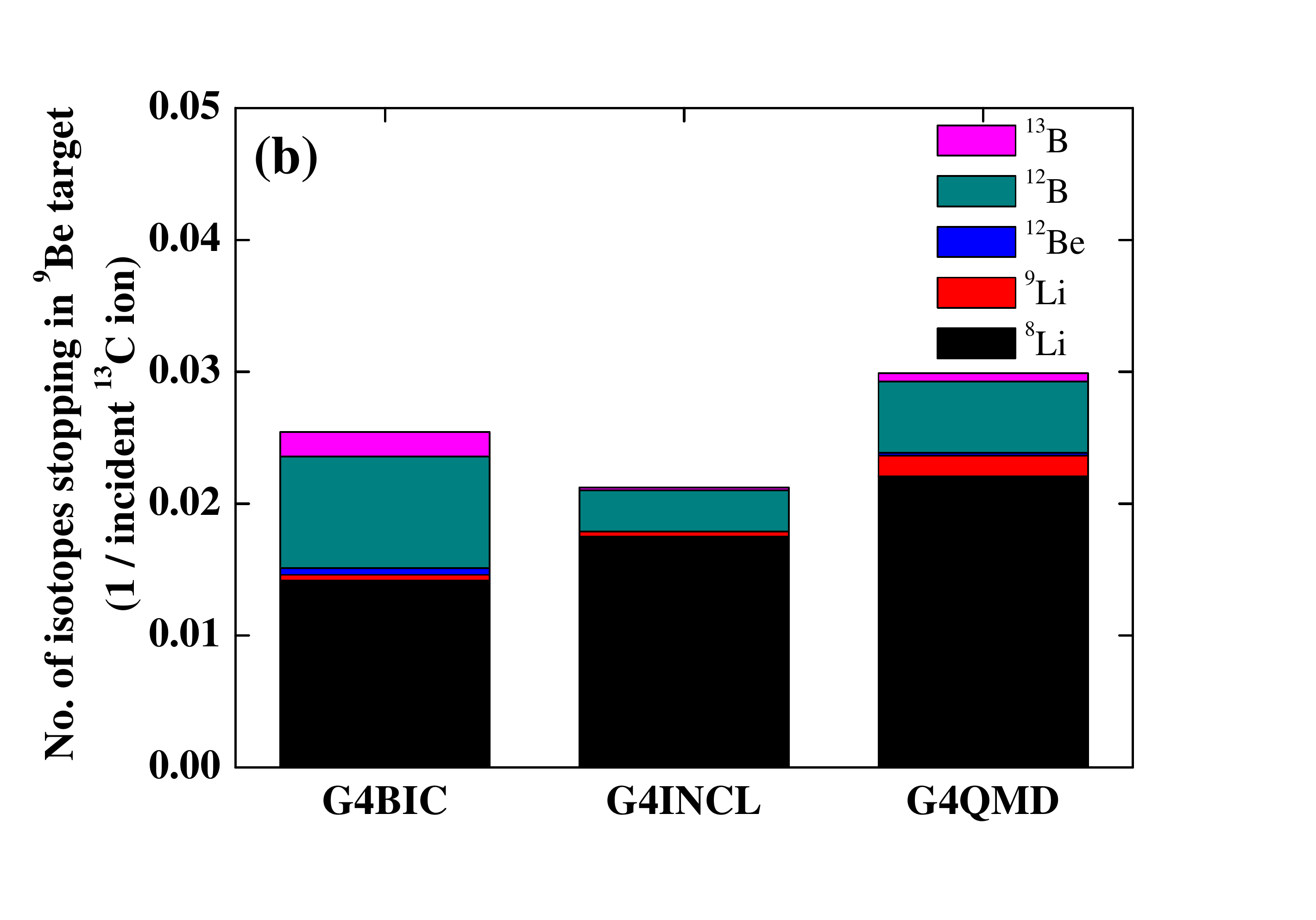, width=3.in}
\caption{(Color online)
Production isotope yields (a) and their sums (b) in $^{9}$Be target obtained from G4BIC, G4INCL, and G4QMD.
$^{6}$He is not shown in (b)
because the neutrino energy
from $^{6}$He is lower than the detection threshold energy of 4 MeV.
}
\label{yieldInBe}
\end{figure}

\subsection{Production isotope yields in $^{9}$Be target}

The numbers of unstable isotopes accumulated
inside the $^{9}$Be target,
obtained using G4BIC, G4INCL, and G4QMD, are plotted in figure \ref{yieldInBe} (a).
The figure shows that
electron antineutrinos via the 75 MeV/u $^{13}$C + $^{9}$Be reaction
are generated through $\beta$ decay from $^{6}$He, $^{8}$Li, $^{9}$Li, $^{12}$Be, $^{12}$B,
and $^{13}$B, where the summation of their fractions are
99\%, 96\%, and 97.0\% for
G4BIC, G4INCL, and G4QMD, respectively.
Because the half-lives of those isotopes are shorter than a few seconds,
the electron antineutrinos are predominantly
produced from the $^{9}$Be target
during the $^{13}$C beam irradiation.

\begin{table}
\caption{Radioactive decay data for the isotopes produced in $^{9}$Be target.}
\begin{tabular}{c|ccccc}
\hline
  Isotope         & half-life & decay mode & Branching ratio & $Q$ Value &\\
                  & [sec.]    &           &      [\%]       & [MeV]  &\\ \hline
  $^{6}$He        & 0.8067   & $^{6}$He $\to$ e$^{-}$ + ${\bar{\nu}}_{e}$ + $^{6}$Li & 100 & 3.5078 &\\ \hline
  $^{8}$Li        & 0.838    & $^{8}$Li $\to$ e$^{-}$ + ${\bar{\nu}}_{e}$ + $^{8}$Be & 100 & 16.097 &\\ \hline
  $^{9}$Li        & 0.1783   & $^{9}$Li $\to$ e$^{-}$ + ${\bar{\nu}}_{e}$ + $^{9}$Be & 50.5 & 13.606 &\\
                  & 0.1783   & $^{9}$Li $\to$ e$^{-}$ + ${\bar{\nu}}_{e}$ + n + $^{8}$Be & 49.5 & 11.941 &\\ \hline
  $^{12}$Be       & 0.0213   & $^{12}$Be $\to$ e$^{-}$ + ${\bar{\nu}}_{e}$ + $^{12}$B & 100 & 11.708 &\\ \hline
  $^{12}$B        & 0.0202   & $^{12}$B $\to$ e$^{-}$ + ${\bar{\nu}}_{e}$ + $^{12}$C & 100 & 13.3689 &\\ \hline
  $^{13}$B        & 0.01736  & $^{13}$B $\to$ e$^{-}$ + ${\bar{\nu}}_{e}$ + $^{13}$C & 100 & 13.4372 &\\
\hline
\end{tabular}
\label{table_1}
\end{table}

\begin{figure}[tbp]
\begin{center}
\epsfig{file=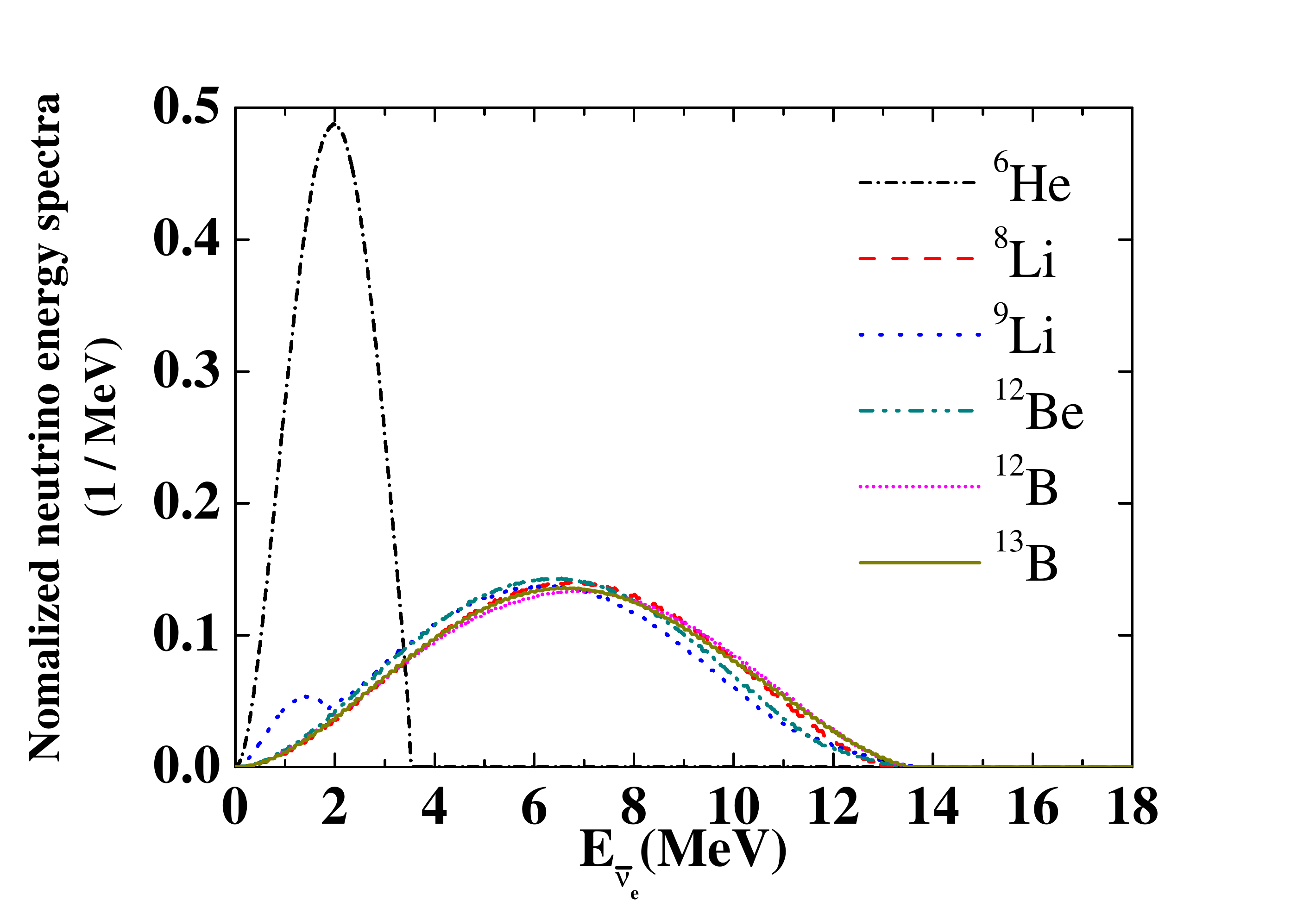, width=4.in}
\end{center}
\caption{(Color online)
Expected neutrino energy distributions from
$^{6}$He, $^{8}$Li, $^{9}$Li, $^{12}$Be, $^{12}$B, and $^{13}$B isotopes.
Decays of 10$^{9}$ for each isotope are simulated,
but all presented results except for $^{12}$Be are divided by 10$^{9}$.
For $^{12}$Be, the results are divided by 2 $\times$ 10$^{9}$
because $^{12}$B which is the daughter nucleus of $^{12}$Be also emits neutrinos.
}
\label{nuEinBe}
\end{figure}

The isotopes except for $^{6}$He
emit electron antineutrinos which have
a maximum flux in the energy range of 6 MeV $<$ $E_{{\bar{\nu}}_{e}}$ $<$ 7 MeV
as shown in figure \ref{nuEinBe},
and their fractions are shown in figure \ref{yieldInBe} (b).
Besides low energy neutrinos from $^6$He, dominant contribution comes from $^{8}$Li and
sub-dominant contribution is $^{12}$B.
The summation of their fractions calculated using G4BIC, G4INCL, and G4QMD
are 0.025, 0.021, and 0.03 per incident $^{13}$C ion, respectively.
This result indicates
very similar shape of neutrino spectra
regardless of the AA reaction model considered in this work.
In the $^{9}$Be target,
$^{10}$Be and $^{14}$C are produced
in addition to the isotopes in figure \ref{yieldInBe} (a).
Due to very long half-lives of
$^{10}$Be (1.5 $\times$ 10$^{6}$ y) and
$^{14}$C (5.7 $\times$ 10$^{3}$ y),
their contributions for neutrino production
are marginal,
and thus they are neglected in this work.

\subsection{Neutrino energy spectra from $^{9}$Be target}

To see expected energy spectrum of all electron antineutrino
from the $^{9}$Be target,
we first obtain the energy spectra from main isotopes
such as $^{6}$He, $^{8}$Li, $^{9}$Li, $^{12}$Be, $^{12}$B, and $^{13}$B.
Radioactive decay data of these isotopes are
tabulated in Table \ref{table_1},
and their neutrino energy distributions are
plotted in figure \ref{nuEinBe}.
The spectrum of the
electron antineutrinos from $^{6}$He have
a peak at $E_{{\bar{\nu}}_{e}}$ = 2 MeV
and a width of 3.5 MeV.
The other spectra
have similar distributions.

\begin{figure}[tbp]
\begin{center}
\epsfig{file=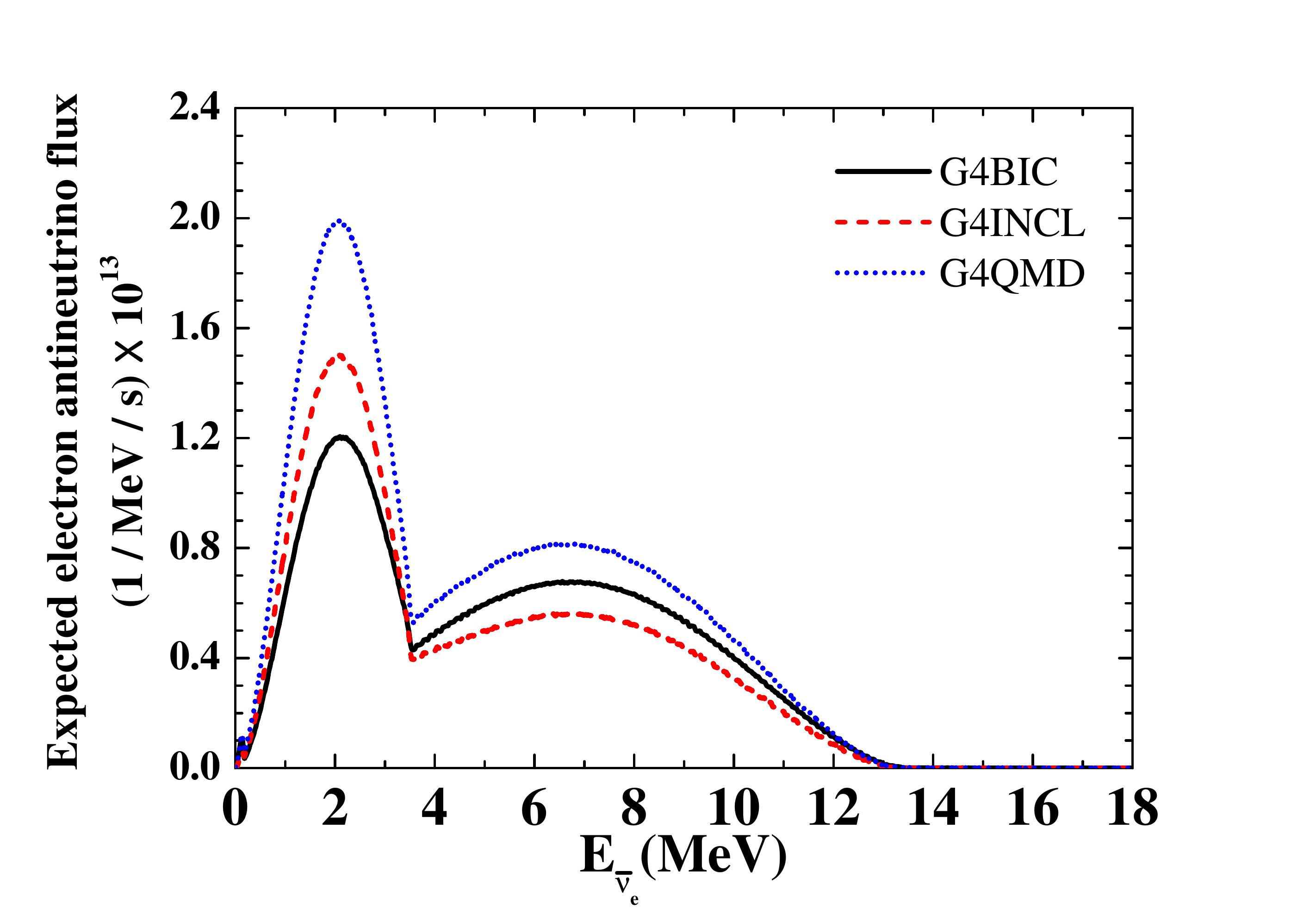, width=4.in}
\end{center}
\caption{(Color online)
Expected electron antineutrino flux in $^{9}$Be target
where the neutrinos are produced by
75 MeV/u $^{13}$C + $^{9}$Be reaction.
The solid line, the dashed line, and the dotted line are
denote the results obtained from G4BIC, G4INCL, and G4QMD,
respectively.
The 300 p$\mu$A current
is assumed.
}
\label{annuEflux}
\end{figure}

Expected electron antineutrino spectra for
the $^{9}$Be target
due to the $^{13}$C + $^{9}$Be reaction
are plotted in figure \ref{annuEflux}, where 75 MeV/u $^{13}$C beams
with 300 p$\mu$A current
are assumed.
Two distinct broad peaks around
$E_{{\bar{\nu}}_{e}}$ = 2 MeV and 7 MeV
are shown in the figure \ref{annuEflux} 
regardless of the nuclear reaction model predictions.
The lower energy peak predominantly come from $^{6}$He
and the higher energy peak originates from
$^{8}$Li, $^{9}$Li, $^{12}$Be, $^{12}$B, and $^{13}$B, as shown in figure \ref{nuEinBe}.
It should be noted that
the spectral shapes in the energy region of E$_{{\bar{\nu}}_{e}} > $ 4 MeV
turn out to be almost identical by multiplying the flux from G4BIC, G4INCL, and G4QMD
by 1.2, 1.46, and 1, respectively.
This characteristic enables us to perform model-independent
shape analysis.

\subsection{Electron antineutrinos produced in the neutrino production targetry}

\begin{figure}[tbp] 
\centering
\epsfig{file=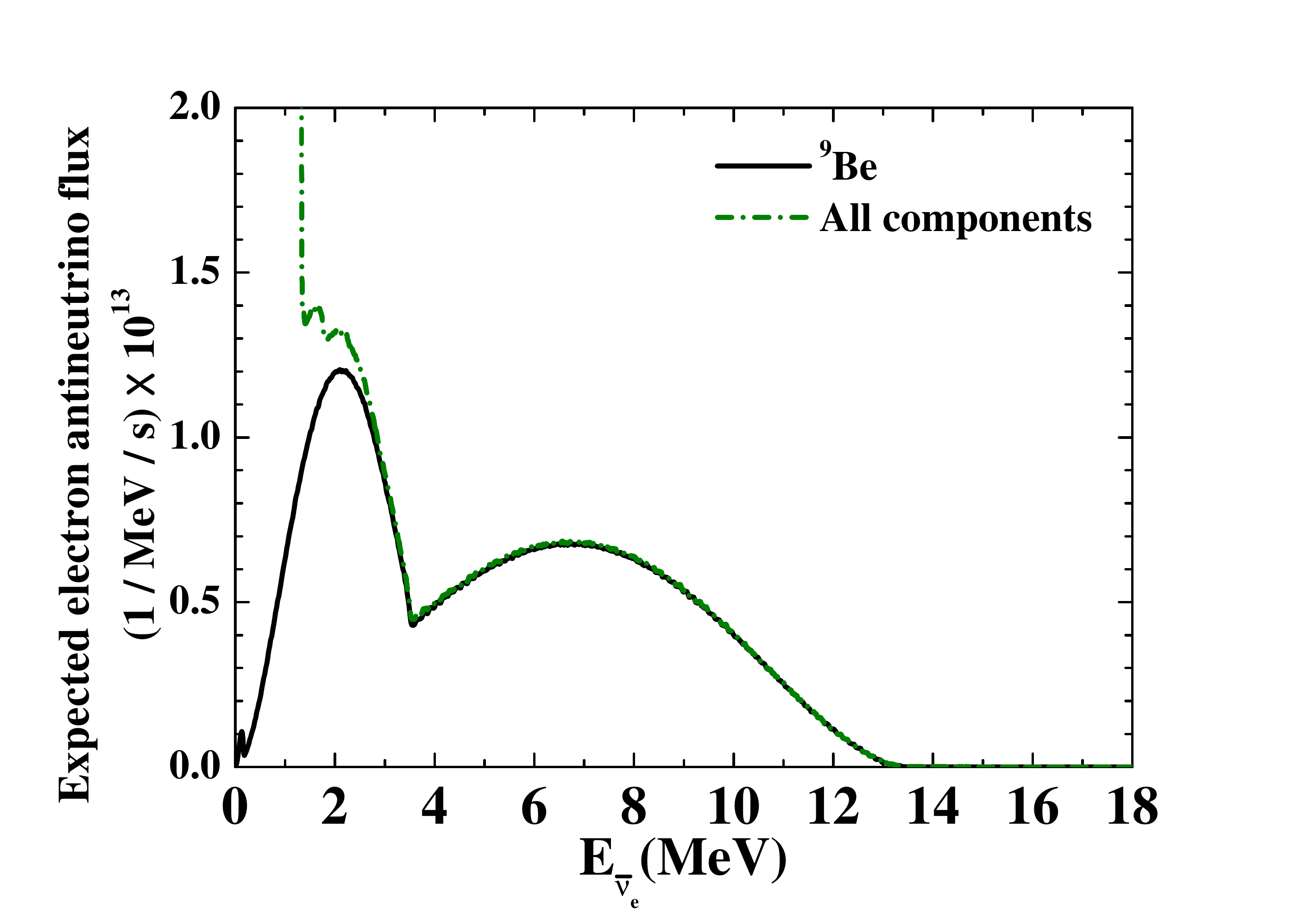, width=4.in}
\caption{(Color online)
Expected electron antineutrino flux produced in the $^{9}$Be target and
the summation of all components in figure \ref{fig1}.
}
\label{figFluxAB}
\end{figure}

Electron antineutrinos can be practically produced in all
components of the production targetry
around the $^{9}$Be target in figure \ref{fig1}.
Figure \ref{figFluxAB} show the calculated results
where the dotted line and the solid line
mean summations of the all components
and the only $^{9}$Be target
in figure \ref{fig1}, respectively.
For decay modes,
we consider fully the decays of the isotopes with
half-life shorter than 10 yr.
Because all of the results from the three different reaction
models mentioned in section \ref{meth-sec}
provide the similar spectra,
we only plot the result using G4BIC in figure \ref{figFluxAB}.

The contribution of the neutrino from the $^{9}$Be target is dominant
in the energy region of E$_{{\bar{\nu}}_{e}} > $ 3 MeV,
but the other components
are dominant for E$_{{\bar{\nu}}_{e}} < $ 1 MeV.
Especially,
in the energy region of E$_{{\bar{\nu}}_{e}} > $ 4 MeV,
the contribution from the $^{9}$Be target is 99\%
of the total neutrinos.
If we consider the neutrino energy cut of E$_{{\bar{\nu}}_{e}}$ = 4 MeV
for electron antineutrino detection,
we can effectively remove background neutrino signals from the other components.
It means that the neutrino source can be concentrated on a small volume of the $^{9}$Be target.
This feature can provide point like source experiments.


\section{Short-baseline electron antineutrino disappearance studies and sterile neutrino
\label{sterile-sec}}


To check the existence of the fourth neutrino,
we compare the event rates under electron antineutrino survival probabilities
in the standard 3-flavor neutrino model (P$_{3}$) and the 3+1 model (P$_{3+1}$).
For the calculation of P$_{3}$,
we use
the equation
given by \cite{PeeRef1}
\begin{eqnarray}
P_{3}
= 1 - {\rm{sin}}^{2} 2 \theta_{13} S_{23} - c^{4}_{13}{\rm{sin}}^{2} 2 \theta_{12} S_{12},
\label{eq:Pee_1}
\end{eqnarray}
where
$S_{23} = {\rm{sin}}^{2}(\Delta m^{2}_{32} L / 4 E)$ and
$S_{12} = {\rm{sin}}^{2}(\Delta m^{2}_{21} L / 4 E)$.
L and E mean the source to detector distance and
the neutrino energy, respectively.
The neutrino oscillation parameters
in Eq.~(\ref{eq:Pee_1})
are taken from a global fit \cite{nu3active}.
Electron-antineutrino survival probabilities
in the 3+1 model can be written as
%
\begin{eqnarray}
P_{\rm{3+1}}
= 1 - 4|U_{e4}|^{2}(1-|U_{e4}|^{2}){\rm{sin}}^{2}( \frac{\Delta m^{2}_{41} L}{4E}).
\label{eq:P_31m}
\end{eqnarray}
The oscillation parameters in Eq.~(\ref{eq:P_31m})
are taken from the best-fit points by the 3+1 model
for the combined short base lines (SBL) and IceCube data sets \cite{s3p1Para}.

\begin{figure}[tbp]
\begin{center}
\epsfig{file=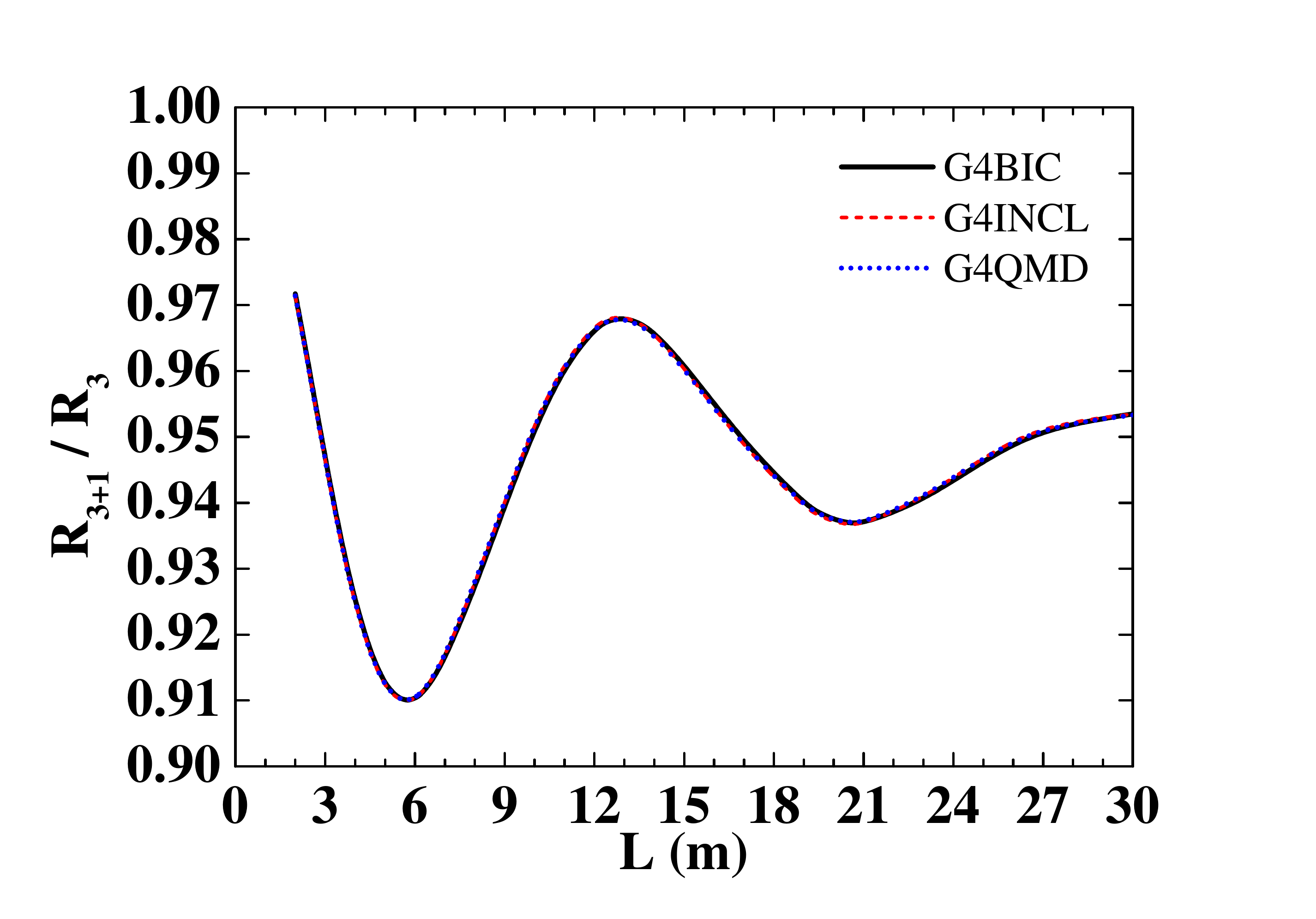, width=4.in}
\end{center}
\caption{(Color online)
Expected event rates with P$_{3+1}$ model
to those with P$_{3}$ model ratios
with respect to L are plotted.
L is the distance between the neutrino source
and the center of the scintillator detector.
}
\label{nuERtot}
\end{figure}

Ratios of the expected total event rate
for the P$_{3+1}$ model
to that for the P$_{3}$ model (R$_{3+1}$-to-R$_{3}$)
are calculated using the G4BIC, G4INCL, and G4QMD models
(see figure \ref{nuERtot}).
Here we assume the neutrino cut-off energy of 4 MeV.
Even for the case without the energy cut,
we obtained almost same results compared to those from calculations with the cut.
All of the results using the three models
are almost unanimous regardless of L.
%
The R$_{3+1}$-to-R$_{3}$ ratios
show oscillation, and
the minimum and the maximum values
of the R$_{3+1}$-to-R$_{3}$
are 0.91 at L = 6 m
and 0.97 at L = 13 m,
respectively.
If there are no sterile neutrinos,
the ratio should be 1.


%
Spectral shapes of the measured neutrinos would also give a chance
to study the existence of fourth neutrinos.\footnote{\protect
The visible energy (E$_{vis}$) of the prompt signal
due to a
positron ($e^{+}$)
is strongly correlated with
the energy of ${\bar{\nu}}_{e}$ (E$_{{\bar{\nu}}_{e}}$),
E$_{{\bar{\nu}}_{e}}$ $\simeq$ E$_{vis}$ + 0.78 MeV, which means that ${\bar{\nu}}_{e}$ energy spectra can be reconstructed
using the E$_{vis}$.
}
To see the effect of possible sterile neutrinos
using the shape analysis like
Daya Bay \cite{ReactBump_Day}, Double Chooz \cite{ReactBump_DOU1},
and RENO \cite{ReactBump_RENO1, ReactBump_RENO2}
experiments,
the event rates are calculated for different L values
using the ring detectors as shown
in figure \ref{fig1} (b).
For the reconstruction of ${\bar{\nu}}_{e}$ energy spectra,
a liquid scintillator detector based on
PROSPECT
type detector \cite{prospect_1}
is considered.
In the present oscillation analysis, 
we assumed 
an statistical error of 1.5\%,
a systematic error of 2\%,
an energy resolution of 4.5{\%}/$\sqrt{\rm{E/MeV}}$,
a position resolution of 15 cm,
and a IBD cross section error of 0.5\% \cite{prospect_1}. 
We use
the cross section obtained using Eq.~(\ref{eq:IBDcs}).

\begin{figure}[tbp]
\begin{center}
\epsfig{file=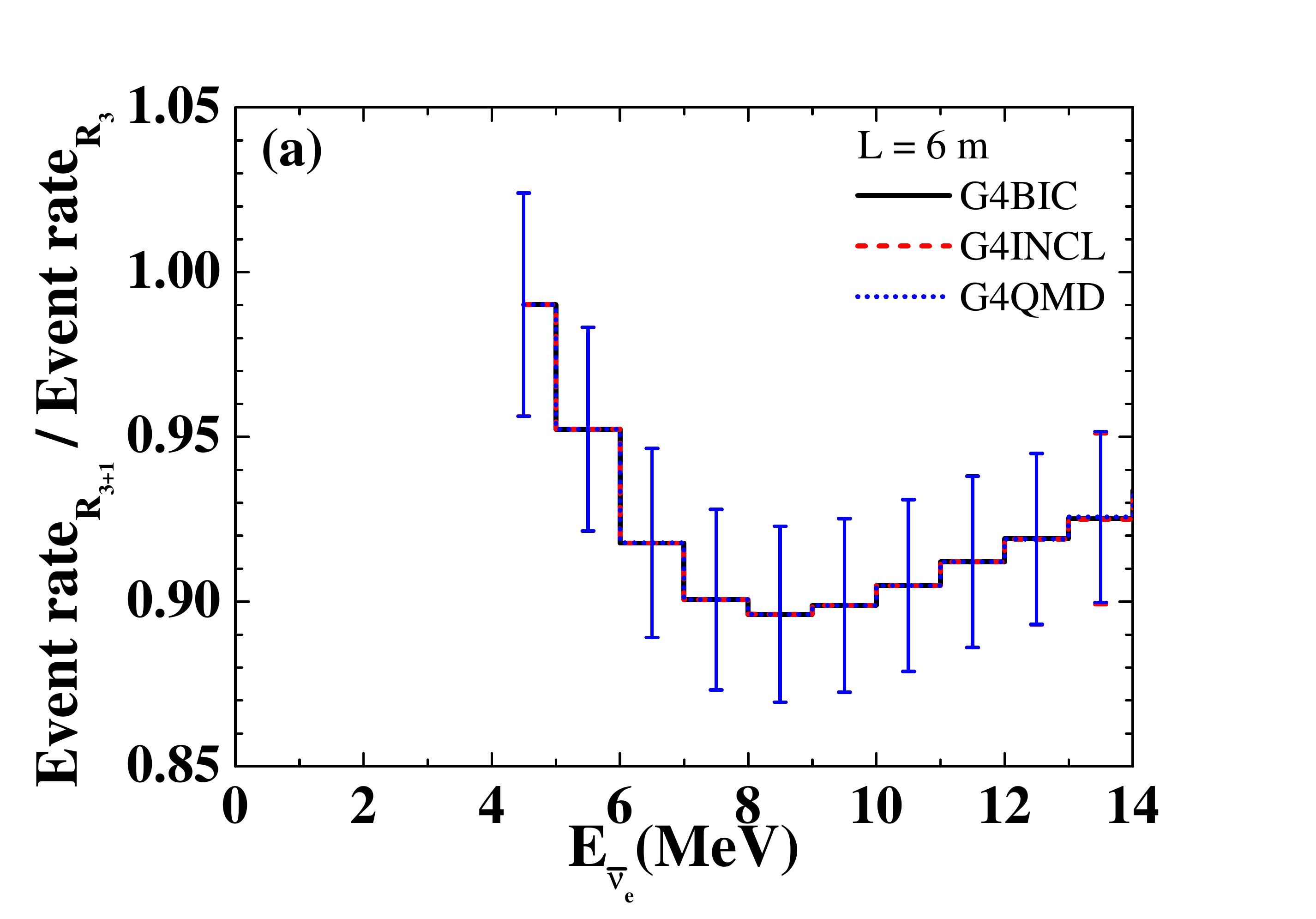, width=3.in}
\epsfig{file=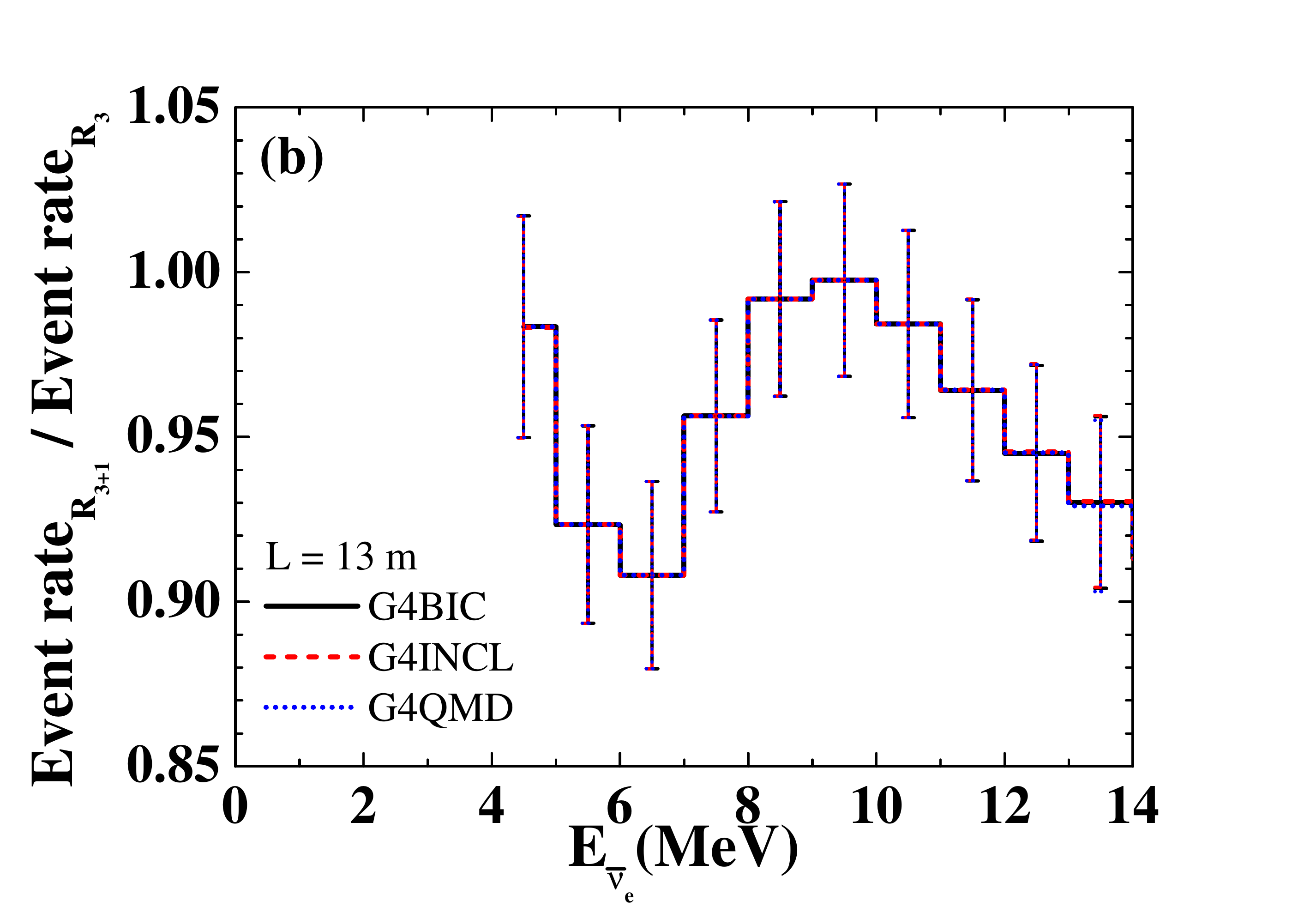, width=3.in}
\end{center}
\caption{(Color online)
Ratios of R$_{3+1}$-to-R$_{3}$
with respect to E$_{{\bar{\nu}}_{e}}$ at (a) L = 6 m and (b) L = 13 m
are shown.
}
\label{nuEdR}
\end{figure}

The R$_{3+1}$-to-R$_{3}$ ratios at L = 6 m and
L = 13 m are plotted in figure \ref{nuEdR}.
It should be also noted that the results in figure \ref{nuEdR}
are almost identical independently of the hadronic models
used in this work.
In the energy region of
4 MeV $<$ E$_{{\bar{\nu}}_{e}}$ $<$ 9 MeV
corresponding to the region of 3 MeV $<$ E$_{vis}$ $<$ 8 MeV, 
the R$_{3+1}$/R$_{3}$ ratio at L = 6 m 
rapidly decreases up to 0.9
as E$_{{\bar{\nu}}_{e}}$ increases.
At the energies of E$_{{\bar{\nu}}_{e}}$ $>$ 9 MeV, however,
the ratio increases as E$_{{\bar{\nu}}_{e}}$ increases.
The comparison of the results in figures \ref{nuEdR} (a) and (b)
in the energy region of 5 MeV $<$ E$_{{\bar{\nu}}_{e}}$
can give a meaningful signal for
the existence of hypothetical neutrinos. 
These characteristics are unique features
of the present work due to 
the compact ${\bar{\nu}}_{e}$ source.
%

\begin{figure}[tbp]
\begin{center}
\epsfig{file=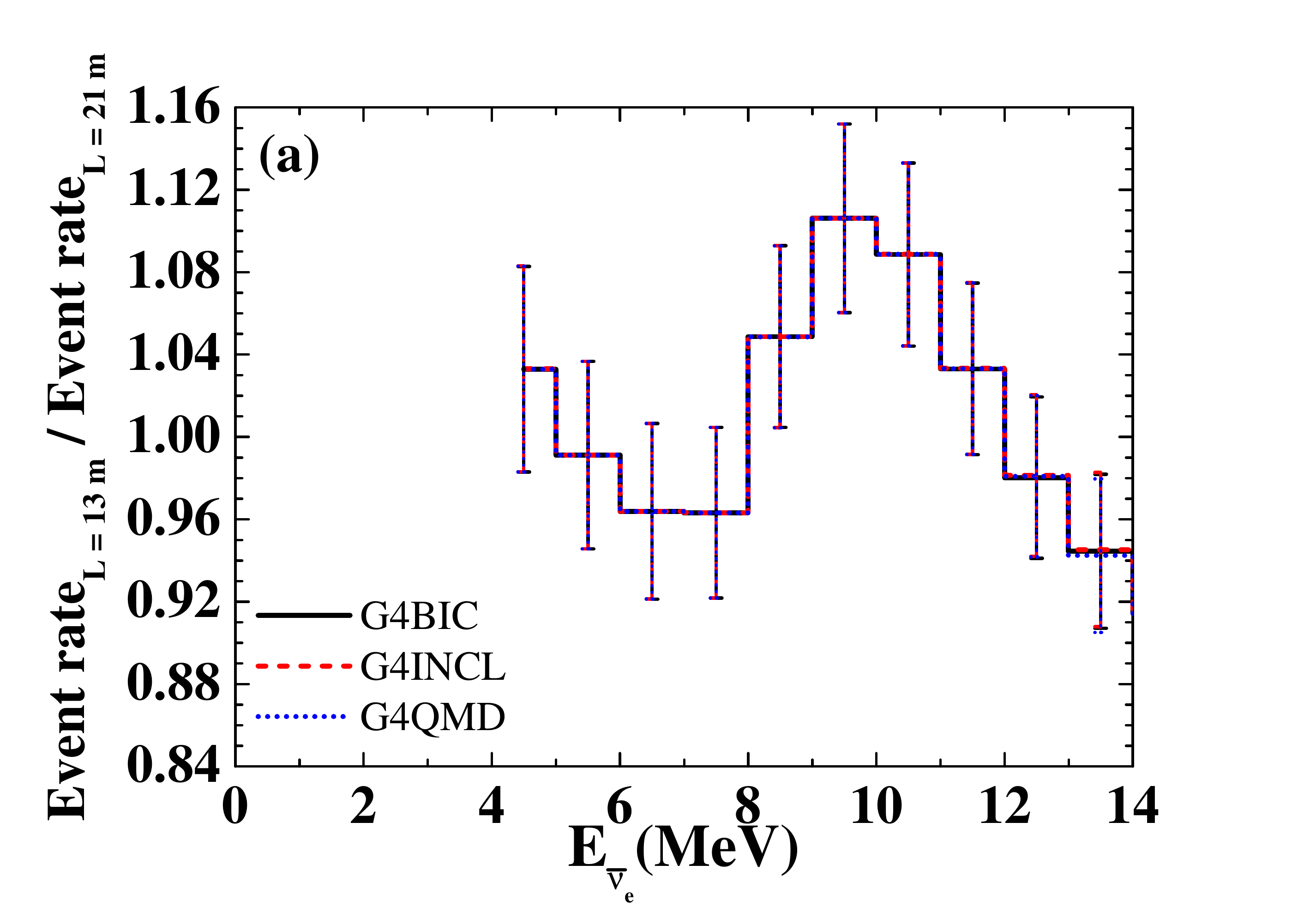, width=3.in}
\epsfig{file=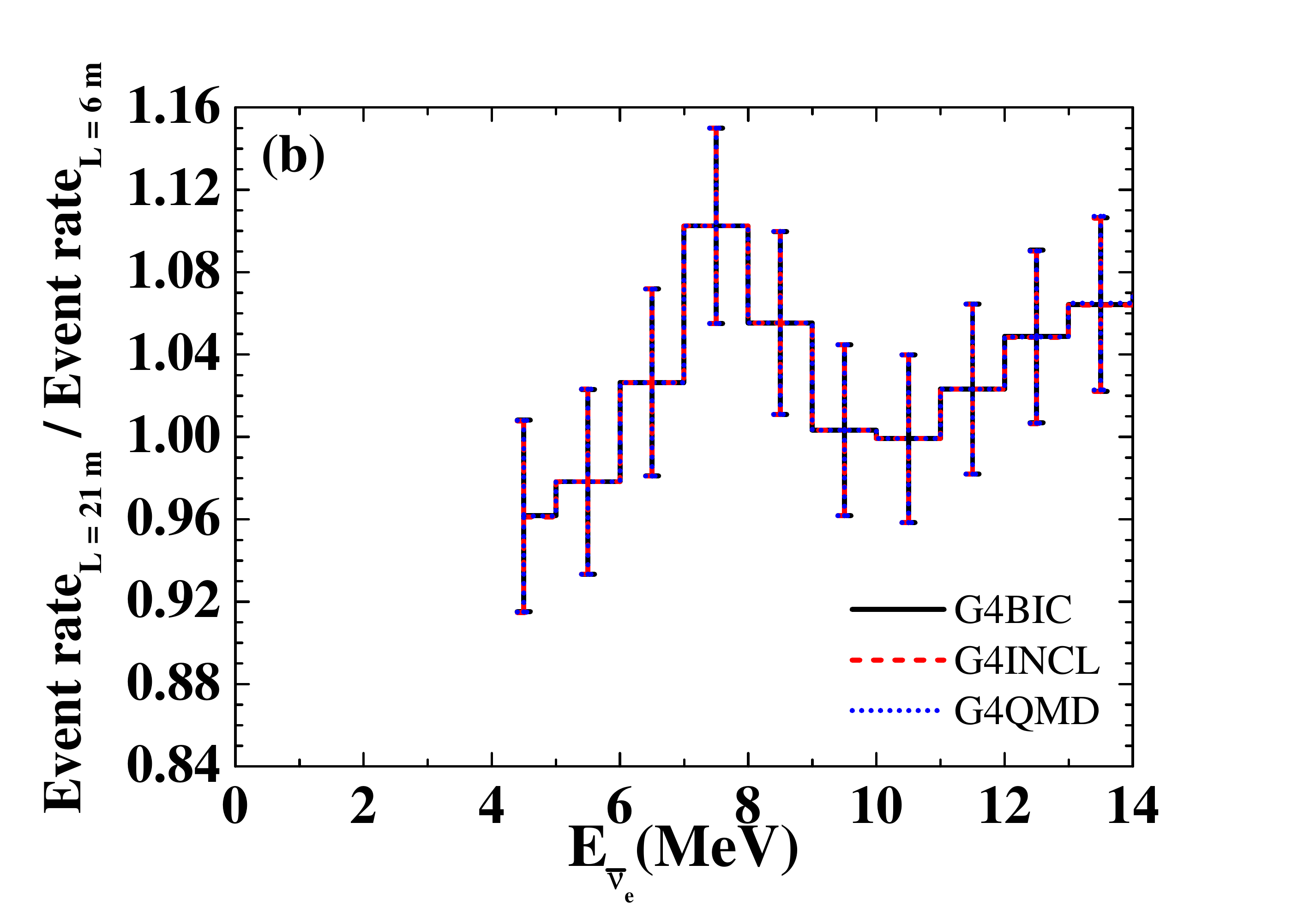, width=3.in}
\end{center}
\caption{(Color online)
Ratios of R$_{3+1}$-to-R$_{3}$
with respect to E$_{{\bar{\nu}}_{e}}$ at different L are shown.
}
\label{nuERratio}
\end{figure}

Also, the comparison among the event rates for different L
(e.g. near and far detectors of reactor neutrino experiments)
can also be possible in this work.
The ratios of the event rate at L = 13 m to that at L = 21 m
with the P$_{3+1}$ model
are plotted in figure \ref{nuERratio} (a).
In the figure,
the maximum and the minimum values are
1.11 at E$_{{\bar{\nu}}_{e}}$ = 9.5 MeV and
0.94 at E$_{{\bar{\nu}}_{e}}$ = 13.5 MeV,
respectively.
If we can measure
approximately 17\% deviation from the
expected events,
we can find a clue to the problem of
whether the P$_{3+1}$
model is the most appropriate scenario.
Figure \ref{nuERratio} (b) shows the results for
R$_{L = 21 m}$/R$_{L = 6 m}$, which gives a different
feature compared to that from figure \ref{nuERratio} (a).
The shapes in figures~\ref{nuERratio}(a) and~\ref{nuERratio}(b)
can give effective chances to search for
the existence of ${\nu}_{s}$
and to test the 3+1 sterile neutrino scenario.

\section{Summary and Conclusion
\label{sum-sec}}

In this work, to investigate the existence of sterile neutrino,
we propose an electron antineutrino source
using $^{13}$C beams based IsoDAR concept
for short-baseline electron antineutrino disappearance study.
The neutrino source is obtained through $\beta^{-}$ decays
of unstable isotopes which are generated from
the $^{13}$C + $^{9}$Be reaction.
Main isotopes for neutrino production
are $^{8}$Li, $^{9}$Li, $^{12}$Be, $^{12}$B, and $^{13}$B.
They have similar half-lives and reaction Q values of $\beta^{-}$ decay,
and thus the neutrino energy spectrum with a single broad peak is expected.

The production yields of those isotopes
are calculated using three different nucleus-nucleus (AA) reaction models.
Even though
different yields of the isotopes are obtained from the models,
the neutrino spectra 
are almost identical. 
This unique feature gives a realistic chance
to neutrino oscillation study through shape analysis, 
regardless of the theoretical AA models considered.

R$_{3+1}$-to-R$_{3}$ ratios at L = 6 m and
L = 13 m
show distinguishable features of the event rates,
and thus it can also give a meaningful signal for
the existence of the hypothetical ${\nu}_{s}$.
Also, complementary comparison studies
among different distance L become feasible.
The expected deviation between the maximum and the minimum
values is approximately 17\%,
and thus it can give an effective
answer of whether
P$_{3+1}$
models is the most appropriate model for the sterile neutrino.

\section*{Acknowledgments}
The work of J. W. Shin is supported by
the National Research Foundation of Korea \ (Grant No. NRF-2015R1C1A1A01054083),
the work of M.-K. Cheoun is supported by
the National Research Foundation of Korea \ (Grant No. NRF-2014R1A2A2A05003548 and NRF-2015K2A9A1A06046598).

\bibliography{mybibfile}

\newpage

\end{document}